\documentclass[10pt]{article}
\usepackage[utf8]{inputenc}
\usepackage[T1]{fontenc}
\usepackage{amsmath}
\usepackage{amsfonts}
\usepackage{amssymb}
\usepackage[version=4]{mhchem}
\usepackage{stmaryrd}
\usepackage{graphicx}
\usepackage[export]{adjustbox}
\graphicspath{ {./images/} }

\title{Fluid Implicit Particle Simulation for CPU and GPU }

\author{Pedro Pereira da Silva Anastácio Centeno\\
(p.silva.centeno@tecnico.ulisboa.pt)\\
\\
João Madeiras Pereira\\
(jap@inesc-id.pt)\\
\\
Instituto Superior Técnico, Lisboa, Portugal}
\date{}

\begin{document}
\maketitle

\begin{abstract}
One of the current challenges in physically-based simulations, and, more specifically, fluid simulations, is to produce visually appealing results at interactive rates, capable of being used in multiple forms of media. In recent times, a lot of effort has been made with regards to this with the use of multi-core architectures, as many of the computations involved in the algorithms for these simulations are very well suited for these architectures. Although there is a considerable amount of works regarding acceleration techniques in this field, there is yet room to further explore and analyze some of them. To investigate this problem, we surveyed the topic of fluid simulations and some of the recent contributions towards this field. Additionally, we implemented two versions of a fluid simulation algorithm, one on the CPU and the other on the GPU using NVIDIA's CUDA framework, with the intent of gaining a better understanding of the effort needed to move these simulations to a multi-core architecture and the performance gains that we get with it.
\end{abstract}


\section*{1. Introduction}
Physically-based fluid animation is an area of great interest in computer graphics and, in particular, can be used to capture stunning visual effects, from clouds and smokes, to fire and explosions, viscous, elastic and plastic flows, sand, droplets, foams, bubbles or storms. In the movie industry, it has its roots in Antz and has been employed in countless other works such as The Lord of the Rings, The Pirates of the Caribbean and The Day After Tomorrow. In the videogame industry it has been used to great effect in games like Shadow of the Tomb Raider.

With the nimble progress we are seeing in computer graphics research, it is now not only possible to perform beautiful fluid simulations, but we can also simulate more advanced effects.

A fluid's behavior can be modeled in multiple ways, ranging from a collection of particles, to a level-set defined surface or a voxelized domain. Each of these, on their own terms, can mold the way the behavior of the fluid is perceived when simulating it. The Fluid Implicit Particle algorithm and its extensions, the current hallmarks for capturing accurate and minute effects, as hybrid methods that combine the advantages of particleand grid-based methods, offer a distinctive approach for handling fluid simulation.

However, this type of simulation involves intensive numerical calculations and, due to the complexity of the employed techniques, we often have to wait several seconds, if not minutes, for every one frame to be computed, which means leaving our computer crunching these scenes overnight or having to wait days for results to appear. These computations can range from simulation steps themselves, in the form of differential operators, integration and particle and grid interpolation, to posterior surface reconstruction algorithms and can even extend to affect the representation of the simulation data.

Fortunately, since FLIP operates on the fact that the potential of the fluid system changes little over the course of a single timestep, each particle can be assumed to be independent of its neighbors. This, in turn, means many of the computations have become more tractable in recent times, as they lend themselves to be trivially parallelizable.

\section*{2. Background}
There is a large body of work on physically-based fluid simulation in computer graphics and, as such, in this section we will only try to touch upon topics of considerable relevance to this project - enough so to help it stand on its own and allow for an outline of most developments that led to it.

\subsection*{2.1. Fluid Simulation}
Physically-based simulation methods model the dynamics of fluids by solving the governing equations, namely the Navier-Stokes equations, which express conservation of momentum, mass and energy for Newtonian fluids. For simplicity's sake, we will focus solely on incompressible and homogeneous fluids.

\subsection*{2.1.1 Navier-Stokes Equations}
There are two of these titular equations: the momentum equation and the incompressibility condition.

The momentum equation is really just Newton's Sec-\\
ond Law of Motion, $\mathbf{F}=m \mathbf{a}$, in disguise. Imagining the fluid as a particle system, where each particle represents a small blob of fluid with mass $m$, volume $V$ and velocity $\boldsymbol{u}$, the same law then tells us how each particle accelerates, allowing us to integrate the system forward in time.

The acceleration term can be written as the material derivative of the particle's velocity, leaving us with $\mathbf{F}=$ $m \frac{D \mathbf{u}}{D t}$. With external forces, which usually amount to just gravity, set aside for now, the remaining forces acting on each particle are due to pressure and viscosity.

For pressure, what matters in the imbalance of higher pressure on one side of the particle than the other, resulting in a force pointing away form the high pressure and towards the low pressure. The simplest way to measure the imbalance at the position of the particle is to take the negative gradient of the pressure which, when integrated by approximation over the particle's volume, results in the pressure force $-V \nabla p$.

For viscosity, there is a force that tries to make a particle move at the average velocity of its neighbours, minimizing the difference in velocity between them. The differential operator that measures how far a quantity is from the average around it is the Laplacian which, when also integrated by approximation of the particle's volume and multiplied by the viscosity coefficient, yields the viscous force $V \mu \nabla^{2} \mathbf{u}$.

Putting it all together, we get a differential equation which dictates how a blob of fluid moves:

$$
m \frac{D \mathbf{u}}{D t}=m \mathbf{g}-V \nabla p+V \mu \nabla^{2} \mathbf{u}
$$

Given that it is erroneous to approximate a fluid to a finite number of particles, we should take the limit as our number of particles goes to infinity and the size of each goes to zero. This poses a problem, as the mass and volume of a particle are then also going to zero. This can be solved by dividing the equation by the volume and only taking the limit subsequently.

Remembering that $m / V$ is the fluid density and $\mu / \rho$ is the kinematic viscosity, this is what we are left with:

\begin{equation*}
\rho \frac{D \mathbf{u}}{D t}=\rho \mathbf{g}-\nabla p+\mu \nabla^{2} \mathbf{u} \Leftrightarrow \frac{D \mathbf{u}}{D t}=\mathbf{g}-\frac{1}{\rho} \nabla p+\nu \nabla^{2} \mathbf{u} \tag{1}
\end{equation*}

Picking an arbitrary chunk of fluid at some instant in time, with volume $\Omega$ and its boundary surface $\partial \Omega$, we can measure how fast said volume is changing by integrating the normal component of its velocity around the boundary.

$$
\frac{d}{d t} \operatorname{Volume}(\Omega)=\iint_{\partial \Omega} \mathbf{u} \cdot \hat{n}
$$

For an incompressible fluid, the volume should then remain constant, i.e. this rate of change is zero:

$$
\iint_{\partial \Omega} \mathbf{u} \cdot \hat{n}=0
$$

Using the divergence theorem, we can translate this to a volume integral and get:

$$
\iiint_{\Omega} \nabla \cdot \mathbf{u}=0
$$

This equation should hold true for any choice of $\Omega$ and the only function that integrates to zero independent of the volume of integration is zero itself, prompting that:

\begin{equation*}
\nabla \cdot \mathbf{u}=0 \tag{2}
\end{equation*}

Known as the incompressibility condition, this is the other part of the incrompressible Navier-Stokes equations.

\subsection*{2.1.2 Lagrangian and Eulerian Viewpoints}
As previously stated, there are multiple approaches to trailing a moving continuum.

On the one hand, much like how we began by deriving the momentum equation, the Lagrangian approach treats a continuum as a particle system - "a set of discrete particles that move through space"[4], each with its own position, velocity, density, mass and other assortment of properties.

On the other hand, the Eulerian approach, instead, "looks at fixed points in space and measures how each fluid quantity changes with time at each of those points"[13]. In Eulerian methods, the Navier-Stokes equations are discretized with grids.

The link between the two viewpoints is the material derivative, which states, for any quantity $q$, that:

$$
\frac{D q}{D t}=\frac{\partial q}{\partial t}+\nabla q \cdot \mathbf{u}
$$

Returning now to equation 1 and applying this result we finally get:

\begin{equation*}
\frac{\partial \mathbf{u}}{\partial t}=-(\mathbf{u} \cdot \nabla) \mathbf{u}+\nu \nabla^{2} \mathbf{u}-\frac{1}{\rho} \nabla p+g \tag{3}
\end{equation*}

\subsection*{2.1.3 Stable Fluids}
At the inception of physically-based fluid simulation, researchers "did not tackle the Navier-Stokes equations directly"[13]. In exchange for simplicity and efficiency, their approahces did not capture a score of interesting phenomena such as overturning waves, sprays and splashes.

In 1999, Jos Stam introduced the first unconditionally stable algorithm that solves the Navier-Stokes equations[12]. Stam achieved this by exercising the idea of time splitting, effectively dividing the Eulerian method in equation 3 into four distinct steps, where $\mathbf{w}_{0}(x)=$ $\mathbf{u}(\mathbf{x}, t)$ and $\mathbf{w}_{4}(x)=\mathbf{u}(\mathbf{x}, t+\Delta t):$

$\mathbf{W}_{0}(\mathbf{x}) \xrightarrow{\text { force }} \mathbf{W}_{1}(\mathbf{x}) \xrightarrow{\text { advect }} \mathbf{W}_{2}(\mathbf{x}) \xrightarrow{\text { diffuse }} \mathbf{W}_{3}(\mathbf{x}) \xrightarrow{\text { project }} \mathbf{W}_{4}(\mathbf{x})$

We can discard the diffusion step because we will mostly be dealing with inviscid fluids.

\subsection*{2.1.4 Hybrid Methods: PIC and FLIP}
Lagrangian techniques avoid continuous re-sampling during the advection step, which allows for finer-grained details, such as turbulent splashing flows, bubbles and foams. Additionally, they have an easier time depicting irregular boundaries. In general, however, this comes at the expense of stability restrictions, numerical smoothing, inability to guarantee incompressibility and a need for complicated re-meshing.

Eulerian techniques, in contrast, allow for larger time steps and, thanks to grid discretizations, have well defined finite differences for computing the projection step. Having said that, they suffer from numerical dissipation in the semi-Lagrangian advection step.

Just as the name proposes, hybrid methods try to combine the advantages of both Lagrangian and Eulerian techniques.

Evans and Harlow presented the Particle in Cell (PIC) method in 1957[7]. The central idea behind this method is that particles represent the fluid domain and carry velocity but, before the projection step, the particles' velocities are transferred to a grid where this step is handled. Once this new velocity field has been computed, the velocites are interpolated back to the particles. This back-and-forth interpolation between particles and grid introduces some numerical dissipation, which leads to fluids appearing more viscous then they are in actuality. Abreu et al. [17] describes the implementation of PIC code in CUDA-enabled GPU.

Brackbill and Rupert later proposed the Fluid Implicit Particle, or FLIP, method in 1986, whose main contributions was computing the change in velocities instead and adding this back to each particle[3]. Effectively, this alternate approach removes half of the smoothing introduced by PIC but introduces some unwanted visible noise. The model solution, then, became to linearly combine both PIC and FLIP, producing close to ideal results. Zhu and Bridson consolidated this with incompressible flow, which became the go-to algorithm for fluid simulation[16].

\subsection*{2.2. Surface Tracking and Reconstruction}
Discretizing and maintaining moving surfaces are essential parts in capturing accurate behavior in physicallybased fluid simulations. There are two main paradigms in which one can categorize the methods employed: implicit (or Eulerian) and explicit (or Lagrangian) surface tracking.

\subsection*{2.2.1 Level-Set Methods}
Unlike other volumetric effects such as smoke or clouds, liquids have clear interfaces and can be plagued by extreme topology changes - "examples include merging, i.e., multiple surfaces becoming one; splitting, i.e., one surface being divided into several; and changes in genus, i.e., opening and closing of holes in the fluid volume"[6]. For this reason, implicit techniques are the tried-and-true approach to describe liquids.

The most common way to define implicit surfaces is through the use of level-set methods, proposed by\\
Zhao et al.[15]. The surface is characterized with a set of distance samples stored in a fixed grid, where it can be interpreted as the zero-contour of a signed distance field, $\Phi(\mathbf{x})$ - positive $\Phi(\mathbf{x})$ indicates $\mathbf{x}$ is outside the fluid region, while negative $\Phi(\mathbf{x})$ identify the inside.

When dealing with implicit surfaces, it is necessary to reconstruct them at render time in order to visualize the fluid domain. One of the first instances of implicit surfaces for particle-based simulations, called blobbies was proposed by Blinn in 1982[1], in which a set of smooth scalar functions (or smoothing kernels) is used to map points in space to a mass density, i.e., each particle of the fluid defines an implicit sphere with a set radius.

In 2005, Bridson and Zhu introduced a method called improved blobbies[16], producing more pleasing results by exactly reconstructing the signed distance field of each isolated particles.

\subsection*{2.2.2 Marching Cubes}
The above methods only extract scalar distance fields from the particles. To convert the field into a triangle mesh, one of the more common and simpler approaches for offline rendering is the marching cubes algorithm, introduced by Lorensen and Cline in 1987[10].

The algorithm is based on evaluating a real-valued function into a discrete scalar field discretized on a regular grid. The standard MC constructs a "facetized isosurface" by processing the signed distance function for every cell corner in the grid in a "sequential, cube-bycube (scanline) manner"[11].

\section*{3. System Overview}
\subsection*{3.1. CPU-based Implementation}
\subsection*{3.1.1 Staggered Grid}
When simulating Eulerian fluids, where grids are used to maintain information, there is a need to not only discretize the simulation in time, but also in space. The most fundamental innovation in this field was the Marker-and-Cell grid, introduced by Harlow and Welch[8]. The MAC grid has a staggered configuration, i.e., it stores different variables at different locations. Figure 1 illustrates a three-dimensional MAC grid.

\begin{center}
\includegraphics[max width=\textwidth]{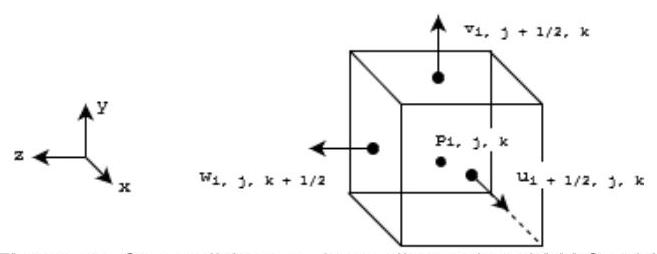}
\end{center}

Figure 1: One cell from a three-dimensional MAC grid.

When applied to FLIP, this means scalars have their sample points located at the center of the cell, whereas sample points of the velocity field (a vector measure) coincide with the cell face boundaries. The benefit of using this configuration comes from the fact that the discretization of differential operators is greatly simplified - namely,\\
we can use accurate central differences for the pressure gradient and for the divergence of the velocity. To estimate the derivative of any quantity $q$ at a grid point $i$, we can do so as:

$$
\left(\frac{\partial q}{\partial x}\right)_{i} \approx \frac{q_{i+1 / 2}-q_{i-1 / 2}}{\Delta \tau}
$$

Where $\Delta \tau$ is the size of a cell. This is "unbiased and accurate to $O\left(\Delta x^{2}\right)^{\prime \prime}[5]$, naturally resulting in smaller-scale errors.

Note also that when manipulating staggered grids, we sample the normal components of the velocity at the center of each of its faces, which intuitively allows us to estimate the amount of fluid flowing into and our of the cell.

Although the MAC grid is suitable for handling pressure and incompressibility, it does come with its downsides. One significant example is when wanting to evaluate the full velocity vector somewhere in the fluid domain, where we always need some sort of interpolation (in the case of three dimensions, trilinearly so) for each component.

Moreover, while half indices, such as $i+1 / 2$, are convenient theoretically and conceptually, an implementation should use integer indices, so we represent an $n$ dimensional MAC grid as a flattened array of cells (see figure 2 as an example).

\begin{center}
\includegraphics[max width=\textwidth]{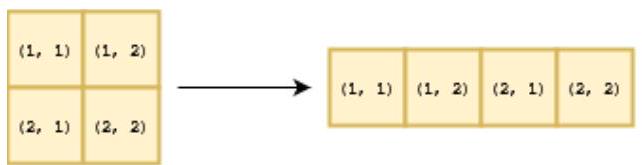}
\end{center}

Figure 2: Representation of a two-dimensional grid as a flattened array.

\subsection*{3.1.2 Seeding}
We handle seeding by defining a sharp region, given by a level set, inside of which the emission density is constant, dropping to zero outside of said region. Inside each cell that represents the fluid domain, the particles are seeded in a jittered grid - much like super-sampling in a renderer - with a number of subcells proportional to a specified density.

\begin{center}
\includegraphics[max width=\textwidth]{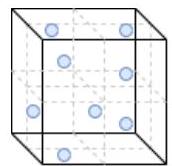}
\end{center}

Figure 3: A density $\rho=2$ totals eight subcells per grid cell, all of which will be populated with one fluid particle randomly located inside.

According to Zhu and Bridson[16], this method helps "avoid aliasing when the flow is underresolved at the simulation resolution". Bridson also proposes that, in order to avoid gaps in the fluid and noise when running the simulation, a fluid cell should contain no less than three and no more than twelve particles[5].\\
At each timestep, if a fluid cell fails to fulfill the specified density, it is naively reseeded until the number of particles it houses reaches the extrapolated value.

\subsection*{3.1.3 Advection}
Advecting the velocity field is akin to moving each particle in the current state of the velocity field. In this step, we apply the physically-motivated semiLagrangian method as introduced by Stam[12]. The Semi-Lagrangian advection traces the particles through the grid, finding the value $q$ at any current grid point by moving back through the path dictated by the velocity field, grabbing the value at the starting point.

Note that at this point in the simulation, the particles are still freely moving, so we're effectively dealing with a Lagrangian measure and, as it happens, in our implementation, we're not carrying any supplementary quantities in a particle, which greatly simplifies the semiLagrangian step. We use a simple forward Euler, which translates applying, for each particle: $\mathbf{x}_{P}=\mathbf{x}_{P}+\mathbf{u} \Delta t$.

This, however, is unconditionally unstable for any timestep, which can cause particles that have penetrated grid boundaries to become stuck. To fix this issue we clamp the particles back into the grid, by moving them in the normal direction of the suspect face by a metric of an arbitrary skin width.

Better results could be obtained by using slightly more sophisticated techniques, such as a higher-order RungeKutta method, but we found that forward Euler yielded adequate results for what we intended.

\subsection*{3.1.4 CFL Condition}
Generally speaking, in physically-based simulations, the goal is to recreate visually pleasing results as fast as feasibly possible. As noted above, our advection scheme is unconditionally unstable, i.e., for any timestep of our choosing, be it bigger or smaller, it can eventually blow up so, in theory, we shouldn't have to bother with stability. Empirically, what we noticed, was that, if we dampenned the rate at which we diverge towards an inaccurate solution by choosing a $\Delta t$ that guarantees CourantFriedrichs-Levy's, or CFL, condition, we got better results.

Bridson notes that, for advection, the true solution is moving at speed $\|\mathbf{u}\|$, so the speed at which numerical information is transmitted must be at least as fast, that is:

$$
\alpha \frac{\Delta \tau}{\Delta t} \geq\|\mathbf{u}\|
$$

Which turns into a condition on the timestep:

$$
\Delta t \leq \alpha \frac{\Delta \tau}{\|\mathbf{u}\|}
$$

Where $\alpha$ is the CFL number, a useful parameter.

Thus, to compute $\Delta t$, we iterate through all the particles to find which is travelling the fastest, and use that in conjunction with a chosen $\alpha=3$ :

\begin{center}
\includegraphics[max width=\textwidth]{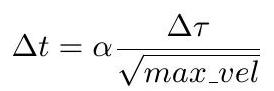}
\end{center}

\subsection*{3.1.5 Particle Binning}
We added auxiliary subdivisions to our domain as a way to hash particles and the cells these occupy in a manner that speeds up grid traversal when necessary, as we get to step over air cells. For this, we created two additional data structures: a vector that stores raw indices to fluid cells in the MAC grid and a grid that bins particles by the cell they are in, as seen in figure 4 :

\begin{center}
\includegraphics[max width=\textwidth]{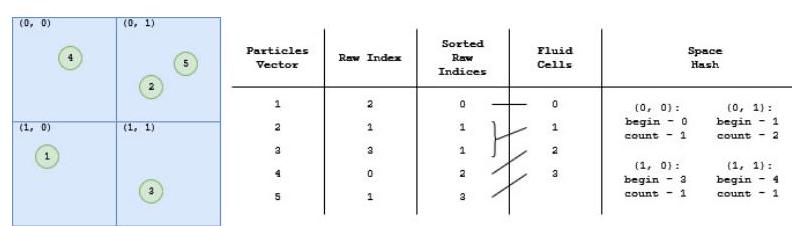}
\end{center}

Figure 4: Representation of the updates carried out for the vector of fluid cells and space hash. First, we compute the raw index for each particle. Then, we sort the vector of particles based on each particle's raw index. Finally, we iterate through the vector, appending a new fluid cell for every new raw index we find and storing the beginning offset and count of particles for each grid cell in the space hash.

The space hash holds a grid-like structure with the same dimensions of the MAC grid and is used during the advection step to survey which cells can be marked as containing fluid versus air (i.e., if their particle count is greater than zero) and the reseeding step where, if the particle count of a cell either exceeds or is below the predefined thresholds, we remove or emit new particles, respectively. The fluid cells vector is used in the projection step.

\subsection*{3.1.6 Particle-to-Grid Transfer}
Bridson[5] mentions the simplest way to compute the total energy of the velocity reaching a certain grid point is to sum, at each grid point, the particle values modulated by a kernel function, $k$, that only gives weight to nearby particles. The kernel function should be adapted to the grid spacing, just as in rendering - if its support is less than $\Delta \tau$ then some particles between grid points may momentarily vanish from the grid (i.e., not contribute to any grid points), but if the support is too much larger than $\Delta \tau$, the method becomes inefficient and it will blur away a lot of the desirable sharpness. He recommends the use of a trilinear hat function, as seen in equation 4.

\begin{gather*}
k(x, y, z)=h\left(\frac{x}{\Delta \tau}\right) h\left(\frac{y}{\Delta \tau}\right) h\left(\frac{z}{\Delta \tau}\right)  \tag{4a}\\
h(r)= \begin{cases}1-r, & 0 \leq r \leq 1 \\
1+r, & -1 \leq r<0 \\
0, & \text { otherwise }\end{cases} \tag{4b}
\end{gather*}

Calculating velocity is done component-wise, and, for example, the $u$ component can be computed as follows:

$$
u_{i+1 / 2, j, k}=\frac{\sum_{P} u_{P} k\left(\mathbf{x}_{P}-\mathbf{x}_{i+1 / 2, j, k}\right)}{\sum_{P} k\left(\mathbf{x}_{P}-\mathbf{x}_{i+1 / 2, j, k}\right)}
$$

With the other components following suit. Bridson further recommends each weight be calculated separately for each grid point due to the non-uniform particle distribution.

Unlike what is suggested in the same chapter, instead of looping through all particles and updating the grid points affected by the current particle, we use what is called a gathering method: we loop over all grid points and calculate the velocity using the previous equation and looping over all particles that affect said grid point, as seen in algorithm ??.

\subsection*{3.1.7 Pressure Solve}
The projection routine will subtract the pressure gradient from the intermediate velocity field computed up to this point,

$$
\mathbf{u}^{n+1}=\mathbf{u}-\Delta t \frac{1}{\rho} \nabla p
$$

so the result satisfies incompressibility inside the fluid, as seen in equationn 2 , and satisfies the solid wall boundary conditions while also respecting the free surface boundary condition that pressure be zero there.

We highlight here the importance of adapting a staggered grid configuration, "making central differences robust". For example, where we need to subtract the $\partial / \partial x$-component of the pressure gradient from the $u$ component of the velocity, there are two pressure values lines up perfectly on either side:

\begin{equation*}
u_{i+1 / 2, j, k}^{n+1}=u_{i+1 / 2, j, k}-\Delta t \frac{1}{\rho} \frac{p_{i+1, j, k}-p_{i, j, k}}{\Delta \tau} \tag{5}
\end{equation*}

This follows suit for the other two components.

As we want our fluid to be incompressible (equation 2), on a grid we approximate this condition with finite differences while requiring that the divergence estimated at each fluid grid cell be zero for $\mathbf{u}^{n+1}$. Remember that the divergence, in three dimensions, is

\begin{equation*}
\nabla \cdot \mathbf{u}=\frac{\partial u}{\partial x}+\frac{\partial v}{\partial y}+\frac{\partial w}{\partial z} \tag{6}
\end{equation*}

which we can also approximate using central differences.

Putting equations 2, 5 and 6 together we reach a numerical approximation to the Poisson problem $-\Delta t / \rho \nabla$. $\nabla p=-\nabla \cdot \mathbf{u}$.

Finally, we can think of this as a large system of linear equations. We can conceptually think of it as a large coefficient matrix, $A$, times a vector consisting of all pressure unknowns, $p$, equal to a vector consisting of the negative divergences in each fluid grid cell, $b: A p=b$.

In our implementation, $p$ and $b$ are grids represented as flattened arrays. With regards to matrix $A$, it is relevant to understand we do not need to store it directly as a matrix. Each row of $A$ corresponds to one equation, i.e., one fluid cell, and its entries are the coefficients of all the pressure unknowns in that equation, almost all of which are zero except (possibly) for the seven entries corresponding to $p_{i, j, k}$ and its six neighbors, $p_{i \pm 1, j, k}, p_{i, j \pm 1, k}$ and $p_{i, j, k \pm 1}$. With this said, not only is $A$ sparse, it is also symmetric, that is, for example, $A_{(i, j, k),(i+1, j, k)}$, the coefficient of $p_{i+1, j, k}$ in the equation for grid cell $(i, j, k)$,\\
has to be equal to $A_{(i+1, j, k}(i, j, k)$ - either it's zero if one of those two cells is not fluid, or it's the same non-zero value. Ergo, we just keep a vector of data for each fluid cell.

The $A$ matrix is a very well-known type of matrix, referred to as the seven-point Laplacian matrix in three dimensions. It is a prototypical sparse matrix and has been meticulously studied so we will not delve very deep into the vast body of work that exists on solving these systems. We implemented a Jacobi, a Gauss-Seidel and a PCG solver on the CPU (of which a more thorough overview can be found in Bridson's titular book[5]).

\subsection*{3.1.8 Grid-to-Particle Transfer}
To finalize a simulation step, the particles need to be updated with the newly computed velocities stored in the MAC grid. This is managed by applying a trilinear interpolation for every velocity component of the eight neighbouring cells of the particle. Essentially, this means each particle will get a weighted velocity depending on its position in a grid cell.

However, since the final velocities are a combination of PIC and FLIP, we first compute the change in velocity and store it and, following this, interpolate the new velocity field and this change based on a blending factor.

\subsection*{3.2. CUDA-based Implementation}
Before introducing our GPU developments, it is important to clarify that the FLIP algorithm, as a whole, was preserved. What we mean by this is that certain steps of the algorithm will not be mentioned in detail in this section as they map directly to our GPU implementation without any non-trivialities.

\subsection*{3.2.1 Data Representation}
We first look at how data for fluid particles and grid nodes should be laid out in GPU memory. Vector fields can be stored in one of two configurations: as an Array of Structures, where each element in the array is an instance of a structure storing vector components of the fluid; or a Structure of Arrays, where each vector component over the entire domain of the fluid is first grouped together and then stored consecutively in an array.

\begin{center}
\includegraphics[max width=\textwidth]{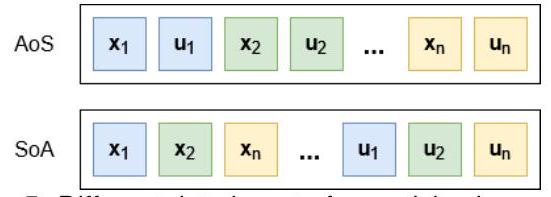}
\end{center}

Figure 5: Different data layouts for particles in memory.

If we were to access each component of a particle concurrently, then AoS would make sense as successive reads of each component would be continuous. However, during the advection step and when binning particles, we assign one thread per particle and not all of the different physical components are accessed at the same time. To put it differently, if we use an AoS configuration, scattered memory access patterns would occur, as each thread would fetch values for each particle, which would be displaced in memory. By adopting a SoA configuration, we can coalesce memory accesses, achieving higher memory bandwidth.

Additionally, we maintain one global array for each velocity component for all grid cells and, for parameters that remain constant throughout the simulation (e.g., gravity accelaration, grid size and spacing, etc.), we utiliza CUDA's constant memory to store them for efficient data access.

\subsection*{3.2.2 Particle Binning}
For the sake of simplicity, all grid cells are the same size spatially, which means each grid cell only has to keep track of the section of the main particle list that it represents and its spatial origin.

The spatial origin of the grid cell is hashed using a 16-bit unsigned-integer, referred to as the cell index, which is used for determining the region of the domain the cell is responsible for and computed as index $=$ $\left(\mathbf{u}_{P}-\right.$ grid\_min $) *$ cell\_size.

In order to compute grid offsets, we do an all-prefixsums operation on the amount of particles in each grid cell. An all-prefix-sums operations, also known as a scan, is described as an operation that "takes a binary associative operator, $\oplus$, with an identity, $I$, and an array of $\mathrm{n}$ elements, $\left[a_{0}, a_{1}, \ldots, a_{n-1}\right]$ and returns the array $\left[I, a_{0},\left(a_{0} \oplus a_{1}\right), \ldots,\left(a_{0} \oplus a_{1} \oplus \ldots \oplus a_{n-2}\right)\right] "[9]$. To achieve a work-efficient parallel scan, we follow the approach described in [9]. In essence, the algorithm is based on the idea of using, conceptually, a balanced binary tree on the input data and sweeping it to and from the root to compute the prefix sum.

Finally, the last step is responsible for moving all particle information from their old locations in memory to the new sorted locations, utilizing the double-buffering technique.

\subsection*{3.2.3 Pressure Solve}
Jacobi solvers are inherently parallelizable but tend to have very slow convergence rates. Meanwhile, conjugate gradient solvers are non-trivial to parallelize. For these reasons, we decided not to develop GPU implementations for either of them. We redirect curious readers to [2] for a fine-grained survey on sparse matrix solvers on the GPU.

Gauss-Seidel utilizes values from the same iteration it is computing, making it also non-trivial to parallelize as is. As it happens, however, for sparse matrices, such as $A$ in our projection step, it is fairly simple to implement a parallel solution. This applies because we can find partitions of variables that can be solved independently. More specifically, if $m_{i, j}$ is non-zero, then the variables $x_{i}$ and $x_{j}$ must not be solved for in parallel; otherwise, it is fair game.

This version of Gauss-Seidel is computed in two interleaved passes, according to the pattern shown in figure 6 . The values of $i$ and $j$ are obtained with the thread

\begin{center}
\includegraphics[max width=\textwidth]{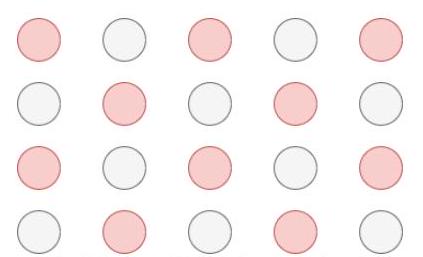}
\end{center}

Figure 6: Gauss-Seidel red-black checkerboard pattern for a 2D grid.

and block information when calling the kernel and fixed, and, for every iteration, we iterate over $k$, updating the solution according to the Gauss-Seidel update rule.

One downside to this approach is that, since matrix values are stored in global memory, it suffers from global memory latency. However, since we store the data for all individual slices of $z$ in a single array, this array's size surpasses that of the allowed shared memory for each thread block, so using shared memory was not an option.

\subsection*{3.3. Surface Reconstruction}
We achieve this with two distinct steps, creating a scalar field out of the simulation data and obtaining vertex data to extrapolate a mesh.

\subsection*{3.3.1 Scalar Field}
For this step, we follow a similar approach to what Zhu and Bridson propose on their paper on improved blobbies[16]. For every grid corner, we calculate the average position and radius of the nearby particles and determine whether that corner is located inside these values. The time complexity for this approach is very inefficient, specially with higher resolution grids, at $O(m \cdot n)$, where $n$ is the number of particles and $m$ the number of grid corners. Using a similar structure to space\_hash described in sec 3.1.5, we can narrow our formula to only the particles we compute a priori to be around the grid corner. What we end up with are negative distances for corners that are close to many particles, and positive distances for those which are not. In effect, this forms a signed distance function of which the definition is the following:

$$
\phi(\mathbf{x})=\left\|\mathbf{x}-\sum_{i} w_{i} \mathbf{x}_{i}\right\|-\sum_{i} w_{i} r_{i}
$$

where $r_{i}$ is the influence radius of a particle. The points are weighted as:

$$
w_{i}=\frac{k\left(\left\|\mathbf{x}-\mathbf{x}_{i}\right\| / R\right)}{\sum_{j} k\left(\left\|\mathbf{x}-\mathbf{x}_{j}\right\| / R\right)}
$$

where $R$ is the search radius of nearby particles to the grid corner and $k$ is the kernel function

$$
k(s)=\max \left(0,\left(1-s^{2}\right)^{3}\right)
$$

which is a function that decays as the distance between particles and the grid corner grows, being exactly zero at a distance equal to the search radius.

\subsection*{3.3.2 Marching Cubes}
We decided to classify each corner as either being below or above the isovalue, giving us 256 possible configurations of corner classification. Two of these are trivial where all points are inside or outside and the cube does not contribute to the isosurface. For all other configurations we need to determine where, along each cube edge, the isosurfaces crosses, and use these edge intersection points to create one or more triangular patches for the isosurface. Each possibility will be characterised by the number of vertices that have values above or below the isosurface: for example, if one vertex is above the isosurface and an adjacent vertex is below the isosurface, then we know the isosurface cuts the edge between these two vertices, the position of the cut being linearly interpolated.

When polygonising a field, we can also control the resolution of the sampling grid (which does not need to match that of the simulation grid). This allows us coarser or finer approximations to the isosurface to be generated, depending on the smoothness we want out of our fluid mesh.

Moreover, for smooth rendering purposes, we also compute normals for each vertex of the triangular faces. To do this, after having created the triangle patches, we average the normals of all the patches that share a vertex, weighted by the inverse of the area of said patch. This makes it so small polygons have greater weight, with the idea being that small polygons may occur in regions of higher surface curvature.

\section*{4. Evaluation}
The test result were captured on a machine with the following specifications: Intel Core i7-8750H @ 2.20GHz CPU, 16GB of DDR4 RAM and an NVIDIA GeForce GTX $1050 \mathrm{Ti}$ GPU.

We compare the performance of our implementations on three different, relatively simple, benchmarks, with varying particle counts and grid sizes. Typical total speedups between our CPU (running the PCG solver) and GPU implementations can be seen in table 1. Every benchmark was run with the specified parameters for 50 full steps. Overall, on the GPU, we can achieve simulation times that are an order of magnitude improved compared to the corresponding CPU simulation, averaging around a speedup of $20 x$.

Table 1: Average time per simulation step. All times are in milliseconds, averaged after 50 steps were taken.

\begin{center}
\begin{tabular}{c|c|c|c|c|c}
\hline
 & \begin{tabular}{c}
Particle \\
$\#$ \\
\end{tabular} & \begin{tabular}{c}
Grid \\
Res \\
\end{tabular} & \begin{tabular}{c}
CPU \\
Time \\
\end{tabular} & \begin{tabular}{c}
GPU \\
Time \\
\end{tabular} & Speedup \\
\hline\hline
Dam-Break & $100 k$ & $32^{3}$ & 451 & 18 & $25 \mathrm{x}$ \\
 & $500 k$ & $64^{3}$ & 2065 & 83 & $24 \mathrm{x}$ \\
 & $1 M$ & $128^{3}$ & 4247 & 204 & $20 \mathrm{x}$ \\
\hline
Double Dam-Break & $100 k$ & $32^{3}$ & 630 & 27 & $23 \mathrm{x}$ \\
 & $500 k$ & $64^{3}$ & 3462 & 154 & $22 \mathrm{x}$ \\
 & $1 M$ & $128^{3}$ & 18114 & 918 & $19 \mathrm{x}$ \\
\hline
Water Drop & $100 k$ & $32^{3}$ & 323 & 16 & $20 \mathrm{x}$ \\
 & $500 k$ & $64^{3}$ & 2960 & 171 & $17 \mathrm{x}$ \\
 & $1 M$ & $128^{3}$ & 16588 & 1056 & $15 \mathrm{x}$ \\
\hline
\end{tabular}
\end{center}

It should be noted that the increasing simulation times with each benchmark are very likely due to the different distribution of particles in space between them: whereas the dam-break example is fairly muted in visual spectacle, the double dam-break and water drop examples result in more splashes and scatters. This, in turn, increases the ratio of "active" grid cells (e.g., grid cells that contain fluid) at a more rapid rate during the initial stages of the simulation, resulting in slower particle binning phases on both approaches.

It is also clear that there is no linearity in simulation times among the samples gathered in table 1 . This can be correlated to the distributions we chose for each of them, i.e., in order to increase the particle count between the dam-break samples, we not only increased the initial particle density per grid cell, but also the measurements (height, width and depth) of the initial domain volumes. This can result in mismatched behavior per sample of each benchmark with, for example, new emerging behavior resulting out of these differences.

Perhaps the most interesting result observed from table 1 is that the speedup decreases with a combination of particle count and grid size. We know we are not allocating enough particles to saturate the GPU, so this speaks to the scalability of our solution. In theory, for smaller particle counts, time spent initializing threads and calling kernels should amount to a sizeable chunk of each timestep, which would become less relevant at higher particle counts and, as such, not impact our speedup as much. This, however, is not our case, as we, again, in general, are met with decreasing speedups.

In order to explore this, we can take a look at figures 7,8 and 9 .

\begin{center}
\includegraphics[max width=\textwidth]{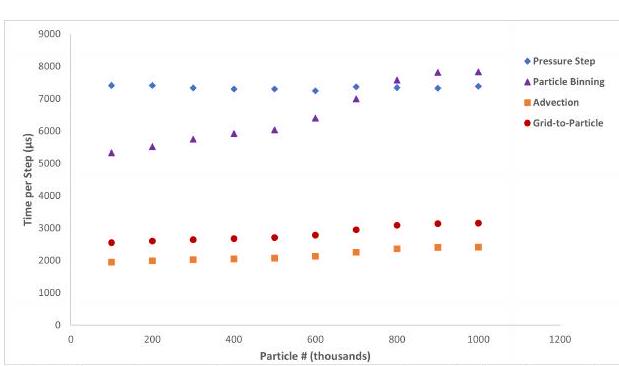}
\end{center}

Figure 7: Time for each stage of the simulation in a simulation step for varying number of particles in a $32 \times 32 \times 32$ grid.

\begin{center}
\includegraphics[max width=\textwidth]{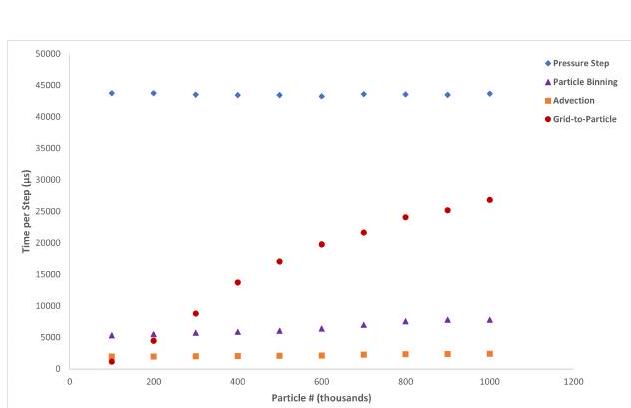}
\end{center}

Figure 8: Time for each stage of the simulation in a simulation step for varying number of particles in a $64 \times 64 \times 64$ grid.\\
For figures 7 and 8 , the simulation was run on the GPU for two standard grid sizes of $32^{3}$ and $64^{3}$, respectively. Straightaway, what jumps out into view is that the pressure solve step takes around the same amount, independently of particle count, for the same grid size. This adds up, given that the size of matrix $A$ is invariable in these cases.

Additionally, for larger grids (e.g., $64^{3}$ ), the pressure step is by far the most expensive for all ranges of particles, whereas for smaller grids (e.g., $32^{3}$ ), the pressure step dominates for fewer than $800 k$ particles, but drops below particle binning and the particle-to-grid transfer (accumulated) for counts larger than that. With this in mind, it is not unreasonable to conclude that what the reason behind this is how most of the data for the pressure step is stored in global memory, which can result in a large amount of memory access conflicts, specially considering the relevant values for each grid cell are in adjacent neighbours which, in turn, neighbour multiple cells. On the other hand, most of the computations for the particle binning step are stored in shared memory and we only access global memory to store the final results, usually in a coalesced manner, which corroborates the values shown.

Another point of interest is the scaling of the particle binning step. Note that, between figures 7 and 8 , it, seemingly, only scales with the particle count and is completely independent of grid size. One might expect to find a variance due to the grid cell size and the distance the particle values have to be moved in memory during this step, but this does not seem to be the case. The same happens for for the advection step, but this is expected as the grid has no bearing in it.

\begin{center}
\includegraphics[max width=\textwidth]{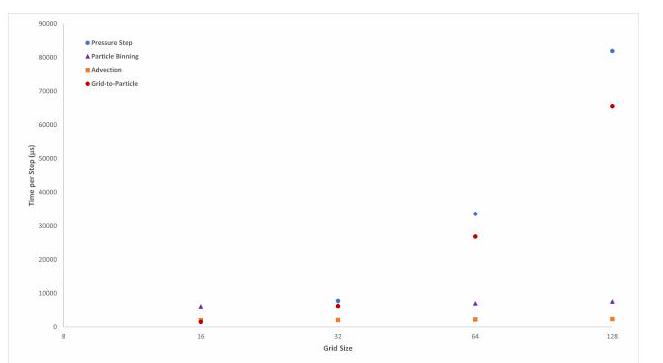}
\end{center}

Figure 9: Time for each stage of the simulation in a simulation step for varying grid side size with $500 k$ particles.

With figure 9, we can take a look at the next parameter of interest, the grid resolution. We ran the simulation with $500 k$ particle for varying resolutions, from $16^{3}$ to $128^{3}$. The pressure step is the primary stage of interest here and, as expected, scales roughly linearly (note that the $\mathrm{x}$-axis is a $\log$ scale) with the number of grid cells. It is also feasible to assume that for grid resolutions that don't saturate the GPU, the pressure step should be bounded by a smaller scaling.

The more subtle curve comes from the grid-to-particle interpolation step, which scales with both parameters, particle count and grid resolution, as can be seen from\\
the above figures combined. The scaling is steeper with grid resolution, which is plausible as particles are sorted and memory access to each of the particles' positions can be coalesced, whereas in order to get face sample values we have to access sporadic memory positions for each grid cell.

In light of these findings, we can understand why the aggregated speedup sees decreasing totals when both parameters are adjusted. We see the pressure step consuming ever more time with higher grid resolutions, which, coupled with an increase in particles, results in more time-intensive particle binning and grid-to-particle interpolation steps as well. From this, it is clear that our solution doesn't scale particularly well. This can be attributed to a lack of use of shared memory, which is only actively employed during binning and, consequently, results in relatively slow memory access patterns. In theory, these could be offset by the use of texture memory to store the diagonal of matrix $A$, as it is only read from, but this is not something we investigated. It is also possible that further tweaks could have been made to kernel parameters, in order to ensure each thread was performing enough work to warrant the cost of its creation.

Before moving on, there is one more thing we want to touch upon by looking at figure 10

\begin{center}
\includegraphics[max width=\textwidth]{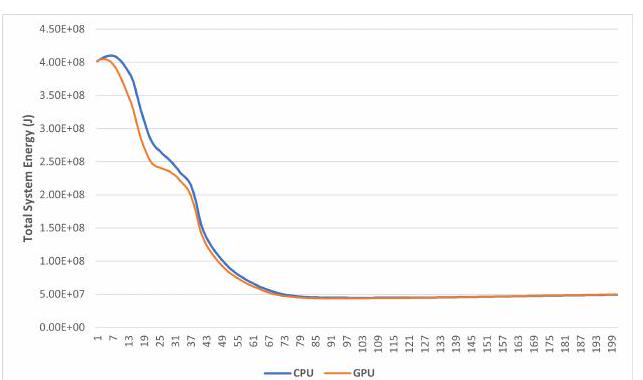}
\end{center}

Figure 10: Total system energy for sets of $25 k$ particles running on the CPU and on the GPU.

This graph shows the variance in system energy for simulations running on the CPU and GPU for 200 full timesteps. We mentioned that our advection scheme, a simple forward Euler, is unconditionally unstable and that can be demonstrated by the downward trajectory in the curve, where the total energy of the system dissipates with each iteration. The samples in figure 10 were taken from the dam-break benchmark, which, as previously indicated, is relatively simple and culminates in a state that allows for less numerical dissipation overall. For more aggressive benchmarks, this scheme can cause the simulation to eventually explode due to the instability introduced.

The gap between the values gathered from both implementation, conversely, can be attributed not to the algorithm itself, but the use of lower precision floating point variables in the GPU implementation, with no care taken to avoid the loss of precision when accumulating intermediate results. A more exact result could be established by performing the calculations at higher preci- sion or by structuring the algorithm in such a fashion that would allow accommodating this difference, but neither option was explored.

\subsection*{4.1. Visual Results}
Table 2: Average time per MC run. All times in milliseconds.

\begin{center}
\begin{tabular}{c|c|c|c|r}
\hline
 & \begin{tabular}{c}
Particle \\
$\#$ \\
\end{tabular} & \begin{tabular}{c}
Grid \\
Res \\
\end{tabular} & \begin{tabular}{c}
Image \\
$\#$ \\
\end{tabular} & \begin{tabular}{r}
MC \\
Time \\
\end{tabular} \\
\hline\hline
Water Drop & $300 k$ & $104^{3}$ & $a)$ & 348 \\
 &  &  & $b)$ & 341 \\
 &  &  & $c)$ & 382 \\
 &  &  & $d)$ & 391 \\
\hline
\end{tabular}
\end{center}

Table 2 shows the run times for $M C$ on two of the benchmarks - the double dam-break and the water drop -, with static particle counts and grid resolution (remember that the grid used for meshing can be of a different resolution than the one used during the simulation), for the images in 11.

The times in table 2 relate to simulations with a number of particles an order of magnitude lesser than what we ideally want to run. Additionally, running with grid resolutions much lower than what we used results in an overall loss of detail, as mentioned in sec ??. As the times in this table indicate, this algorithm is not suited for online rendering, i.e., rasterizing and observing the mesh while the simulation is running in real-time, when running large-scale (this is, with many millions of particles) simulations, since for comparably smaller particle counts we get dangerously close to one second per frame which is not acceptable.

\subsection*{4.2. Discussion}
All in all, our study emphasizes the idea that fluid simulations are highly parallelizable applications. After identifying the steps of the algorithm that were not trivially convertible to CUDA, such as particle binning (which is indispensable for an efficient particle-to-grid transfer) and the projection step, we nevertheless landed on fairly similar approaches between the two implementations, with a high speedup gained from moving to the GPU.

As discussed in the previous section, there are some limitations to our work which could be addressed in future studies. Firstly, the most harmful of these is the lacking use of shared memory or other more efficient memory options at our disposal (e.g., textures). We can see sharp declines in speedup (and, in reverse, inclines in simulation times) due in very big part to the steps where we make the most use out of global memory (such as the pressure solve and grid-to-particle transfer). We believe that approaches that take bigger care when engaging with thread workload and global memory bandwidth could achieve much higher speedups, including for higher particle counts and grid resolutions.

Secondly, although red-black Gauss Seidel is a good midpoint for parallel solvers, we believe that other solvers such as a matrix-free Conjugate Gradient method (as

\begin{center}
\includegraphics[max width=\textwidth]{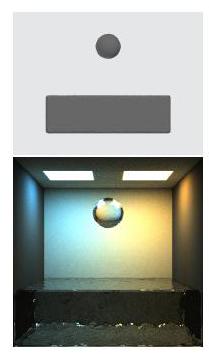}
\end{center}

(a)

\begin{center}
\includegraphics[max width=\textwidth]{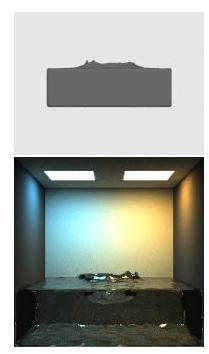}
\end{center}

(b)

\begin{center}
\includegraphics[max width=\textwidth]{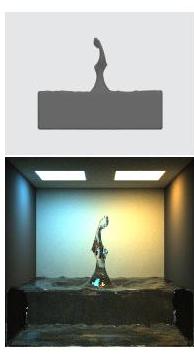}
\end{center}

(c)

\begin{center}
\includegraphics[max width=\textwidth]{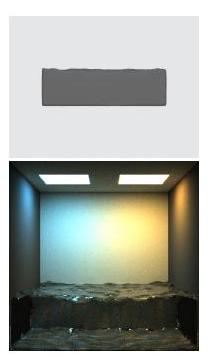}
\end{center}

(d)

Figure 11: Fluid meshes for the water drop benchmark rasterized with Marching Cubes and rendered using PBRTv3 (available here).

seen in [14]) would bring benefits to our implementation, with the potential to not only disregard the use of global memory entirely in the projection step but also introduce methods that, in theory, take fewer iterations to converge.

Finally, another thing of note is that the mindset for our CPU implementation was not one that aimed for effectiveness. Rather, we wanted to get a better grasp of the algorithm and did not concern ourselves with this, which can, in part, have brought about skewed empirical results when compared to our GPU implementation.

\section*{5. Conclusions}
This study was conducted with the goal of getting a better understanding of physically-based simulations, computationally and visually, and how these are impacted with the growing trend of GPGPU. We wanted to come away with concrete conclusions regarding the performance impact of a multi-core approach and feel like our CUDA implementation does just that, allowing for better performance in real-time for simulations with higher particle counts and finer-grained grid resolutions.

To investigate this problem, we started off by devising an implementation of the FLIP algorithm in the CPU and exploring the core steps of said algorithm, to deepen our bases and assimilate what exactly composes the algorithm and how to extract the results we desire in an efficient manner. Following this, we adapted the algorithm to an implementation in CUDA, maintaining the simplicity in the algorithm steps to the best of our ability.

While the end results were what we expected, we learned that a certain care has to be taken into account when implementing the FLIP algorithm on the GPU. Over the course of this project we analyzed the methods we could use to assimilate this implementation, with the main issues we identified being binning particles and the pressure solve step. The conclusions we came to and the details we had to pore over were that: either we use sparse data to continually keep track of only the grid cells that contain fluid, or we need a way to properly sort particles in order to ensure that accumulating values to the grid can be done in a structured manner; and that not all solvers can be trivially implemented in a parallel architecture and we need to opt between better convergence rates or shorter development time.

In the end, however, we cannot say that our work brings relevant contributions to the current landscape of fluid simulations as most of what we worked with has been explored, even to greater detail, in other studies. With this in mind, still, in the end, we achieved what we set out to do, which was to improve our understanding of this specific field and those adjacent to it and showcase that fluid simulations are indeed transparently convertible to multi-core architectures.

\section*{References}
A generalization of algebraic surface drawing. In ACM Transactions on Graphics, page 235-256. ACM, 1982.

[2] J. Bolz, I. Farmer, E. Grinspun, and P. Schröder. Sparse matrix solvers on the gpu: Conjugate gradients and multigrid. SIGGRAPH '03, page 917-924, New York, NY, USA, 2003. Association for Computing Machinery.

[3] J. U. Brackbill and H. M. Ruppel. Flip: A method for adaptively zoned, particlein-cell calculations of fluid flows in two dimensions. In Journal of Computational Physics, page 314-343. Academic Press Professional, Inc., 1986.

[4] C. Braley and A. Sandu. Fluid simulation for computer graphics: A tutorial in grid based and particle based methods. 2009.

[5] R. Bridson. Fluid Simulation for Computer Graphics, Second Edition. Taylor \& Francis, 2015.

[6] T. Brochu. Toward better surface tracking for fluid simulation. In IEEE Computer Graphics and Applications, pages 74-81, 2015.

[7] M. Evans and F. Harlow. The particle-in-cell method for hydrodynamic calculations. Technical report, Los Alamos Scientific Lab., N. Mex., 1957.

[8] F. H. Harlow and J. E. Welch. Numerical calculation of time-dependent viscous incompressible flow of fluid with free surface. In The Physics of Fluids, pages 2182-2189, 1965.

[9] M. Harris, S. Sengupta, and J. Owens. Parallel prefix sum (scan) with CUDA, volume 39, pages $851-.082007$

[10] W. E. Lorensen and H. E. Cline. Marching cubes: A high resolution 3d surface construction algorithm. In SIGGRAPH Comput. Graph., page 163-169. ACM,

[11] T. Newman and H. Yi. A survey of the marching cubes algorithm. Computers \& Graphics, 30:854-879, 102006

[12] J. Stam. Stable fluids. In Proceedings of the 26th annual conference on Computer graphics and interactive techniques, pages 121-128, 1999.

[13] J. Tan and X. Yang. Physically-based fluid animation: A survey. In Science in China Series F: Information Sciences, pages 723-740, 2009.

[14] K. Wu, N. Truong, C. Yuksel, and R. Hoetzlein. Fast fluid simulations with sparse volumes on the gpu. In Computer Graphics Forum, pages 157-167, 2018.

[15] H. Zhao, S. Osher, and R. Fedkiw. Fast surface reconstruction using the level set method. In Proceedings IEEE Workshop on Variational and Level Set Meth ods in Computer Vision, pages 194-201, 2001.

[16] Y. Zhu and R. Bridson. Animating sand as a fluid. In ACM Transactions on Graphics, page 965-972. ACM New York, 2005.

[17] P. Abreu, R. A. Fonseca, J. M. Pereira and L. O. Silva. PIC Codes in New Processors: A Full Relativistic PIC Code in CUDA-Enabled Hardware With Direct Visualization. In IEEE Transactions on Plasma Science, vol. 39, no. 2, pages 675-685, 2011.

\end{document}